\documentclass[preprints,review,accept,moreauthors,pdftex]{mdpi}
\firstpage{1} 
\pubvolume{xx}
\issuenum{1}
\articlenumber{5}
\pubyear{2020}
\copyrightyear{2020}
\history{Rec: 7/04/2020; Acc: 29/04/2020; {\it Galaxies} special issue: {\it Star Formation in the UV}, ed. Jorick Vink. (in press).}

\setitemize{parsep=6pt,itemsep=0pt,leftmargin=*,labelsep=5.5mm}
\setenumerate{parsep=6pt,itemsep=0pt,leftmargin=*,labelsep=5.5mm}
\setlist[description]{itemsep=0mm}  

\Title{On the Mass Accretion Rates of Herbig Ae/Be Stars. Magnetospheric Accretion or Boundary Layer?}

\Author{Ignacio Mendigut\'\i{}a}
\AuthorNames{Ignacio Mendigut\'\i{}a}

\address[1]{%
Centro de Astrobiolog\'{\i}a (CSIC-INTA), Departamento de Astrof\'{\i}sica, ESA-ESAC Campus, 28691 Madrid, Spain; imendigutia@cab.inta-csic.es\\
}

\abstract{Understanding how young stars gain their masses through disk-to-star accretion is of paramount importance in astrophysics. It affects our knowledge about the early stellar evolution, the disk lifetime and dissipation processes, the way the planets form on the smallest scales, or the connection to macroscopic parameters characterizing star-forming regions on the largest ones, among others. In turn, mass accretion rate estimates depend on the accretion paradigm assumed. For low-mass T Tauri stars with strong magnetic fields there is consensus that magnetospheric accretion (MA) is the driving mechanism, but the transfer of mass in massive young stellar objects with weak or negligible magnetic fields probably occurs directly from the disk to the star through a hot boundary layer (BL). The intermediate-mass Herbig Ae/Be (HAeBe) stars bridge the gap between both previous regimes and are still optically visible during the pre-main sequence phase, thus constituting a unique opportunity to test a possible change of accretion mode from MA to BL. This review deals with our estimates of accretion rates in HAeBes, critically discussing the different accretion paradigms. It shows that although mounting evidence supports that MA may extend to late-type HAes but not to early-type HBes, there is not yet a consensus on the validity of this scenario versus the BL one. Based on MA and BL shock modeling, it is argued that the ultraviolet regime could significantly contribute in the future to discriminating between these competing accretion scenarios. }

\keyword{young stars; pre-main sequence objects; T Tauri stars; Herbig Ae/Be stars; protoplanetary disks; accretion disks; magnetic fields; magnetospheric accretion; boundary layer; ultraviolet}

\begin{document}

\section{Introduction}
\label{Sect:intro}

The evolution of young stars comprises several stages from the initial collapse in a molecular cloud until they enter the main sequence (MS), when the central objects reach enough temperature to burn hydrogen~\citep{Lada87}. Although massive young stellar objects (MYSOs, defined from the stellar mass that will lead to a final supernova collapse; M$_*$ $>$ 10 M$_{\odot}$) keep their optically thick envelopes during their fast early evolution, lower-mass stars show an optically visible pre-main sequence (PMS) phase. Optically visible PMS stars are commonly classified based on their stellar masses. The lower-mass objects are the T Tauri stars (TT; 0.1 $<$ M$_*$/M$_{\odot}$ $<$ 2), divided into ``Classical'' (CTT) or ``Weak'' (WTT) depending on whether there are signs of ongoing accretion onto the central star. A stellar mass $\sim$2.5~M$_{\odot}$ and temperature $\sim$8500~K is the rough limit below which the sub-photospheric regions are still fully convective (although convective envelopes are still present up to $\sim$4~M$_{\odot}$) and thus in principle capable of generating magnetic fields through the dynamo process (e.g.,~\citep{Simon02,Villebrun19} and references therein). As we will see in this review, the presence or absence of stellar magnetic fields is fundamental to understand how young stars gain their masses through accretion. The higher mass counterparts of CTTs are the Herbig Ae/Be stars (HAeBe; 2 $<$ M$_*$/M$_{\odot}$ $<$ 10). Essentially, HAeBes are young stars ($\leq$10 Myr) with spectral types A and B, showing H$\alpha$ and other emission lines in their spectra, and a circumstellar disk associated with infrared excess on top of the photospheric emission. The general properties of HAeBes have been discussed in detailed specific reviews~\citep{Perez97,Waters98}, and the reader can consult the online slides, proceedings, and collections associated with more recent conferences devoted to these stars (e.g.,~\citep{deWit14}).\footnote{\url{http://www.eso.org/sci/meetings/2014/haebe2014.html}.}$^,$\footnote{\url{https://starry-project.eu/final-conference/}.}

Several approaches that involve vastly different spatial scales are necessary to understand how material collapses from larger to smaller structures that will finally lead to the formation of the individual stars. However, in last term understanding star formation requires knowing how the circumstellar material actually accretes onto the stellar surface at scales $\ll$ 1 au. In turn, understanding stellar accretion may have implications on the way that ``macroscopic'' parameters like the star formation rate (SFR) are estimated~\citep{Padoan14,Mendi18}, or even on the formation process of planets at the smallest scales~\citep{Lin96}. It is presently accepted that basically all young stars are surrounded by circumstellar disks, and even MYSOs may accrete a non-negligible part of their final masses through these structures (see e.g., the review in~\citep{Beltran16}). Therefore, it is necessary to obtain accurate estimates of the disk-to-star mass accretion rate ($\dot{M}_{\rm acc}$) and thus to know how such an accretion proceeds across a large range of stellar masses. Indeed, deriving $\dot{M}_{\rm acc}$ values requires a formal scenario from which direct observations can be interpreted, and different paradigms can lead to different accretion rate estimates based on the same observational data. For CTTs there is consensus that accretion is magnetically driven according to the magnetospheric accretion scenario (MA~\citep{Uchida85,Konigl91,Shu94}), while for more massive stars without magnetic fields accretion may proceed directly from the disk to the star through a boundary layer (BL~\citep{Lynden74}). In this respect, HAeBes represent a fundamental regime that bridges the gap between the accretion properties of low-mass CTTs and those of MYSOs. Moreover, early-type Herbig Be stars (HBes) are the most massive stars for which direct accretion signatures can still be observed, given that MYSOs embedded in their natal clouds are opaque to robust accretion tracers that emit in the optical and ultraviolet (UV).

This review focuses on our estimates of accretion rates in HAeBe stars, thus discussing the way that disk-to-star accretion may proceed in these sources. Given that much of our current understanding of this topic has been partially inspired by the better known TT stars, Section~\ref{sect:perspective} starts with an historical overview about how accretion has been understood and measured first for these objects and then for the HAeBes. Section~\ref{sect:MA_validity} critically discusses the viability of MA in HAeBes mainly focused on the required and the observed magnetic fields. The different ways to measure accretion rates and the corresponding accuracies based on MA  are described in Section~\ref{sect:MA}. In Section~\ref{sect:BL} the few accretion rate estimates based on the BL scenario that are available for the HAeBe regime are discussed in comparison with the MA measurements. Then it is argued in Section~\ref{sect:uv} that the UV regime may be critical to test the validity of both competing scenarios. Finally, Section~\ref{sect:summary} includes some concluding~remarks. 

\section{A Brief Historical Perspective}
\label{sect:perspective}
Although HAeBes tend to be brighter than TTs, which facilitates the detailed study of some of their disks through high-spatial resolution techniques, TT stars are generally better understood than the intermediate- and high-mass star regimes. Apart from being the precursors of Solar-like stars similar to our own, the main reason is that TTs are comparatively easier to find. This results from the fact that the shape of the initial mass function favors the formation of low-mass objects, and because the PMS phase is shorter as the stellar mass increases. Thus, in many aspects---and concerning accretion in particular---our knowledge of HAeBes is at least partially guided by previous works on~TTs. 

\begin{figure}[H]
\centering
\includegraphics[width=8 cm]{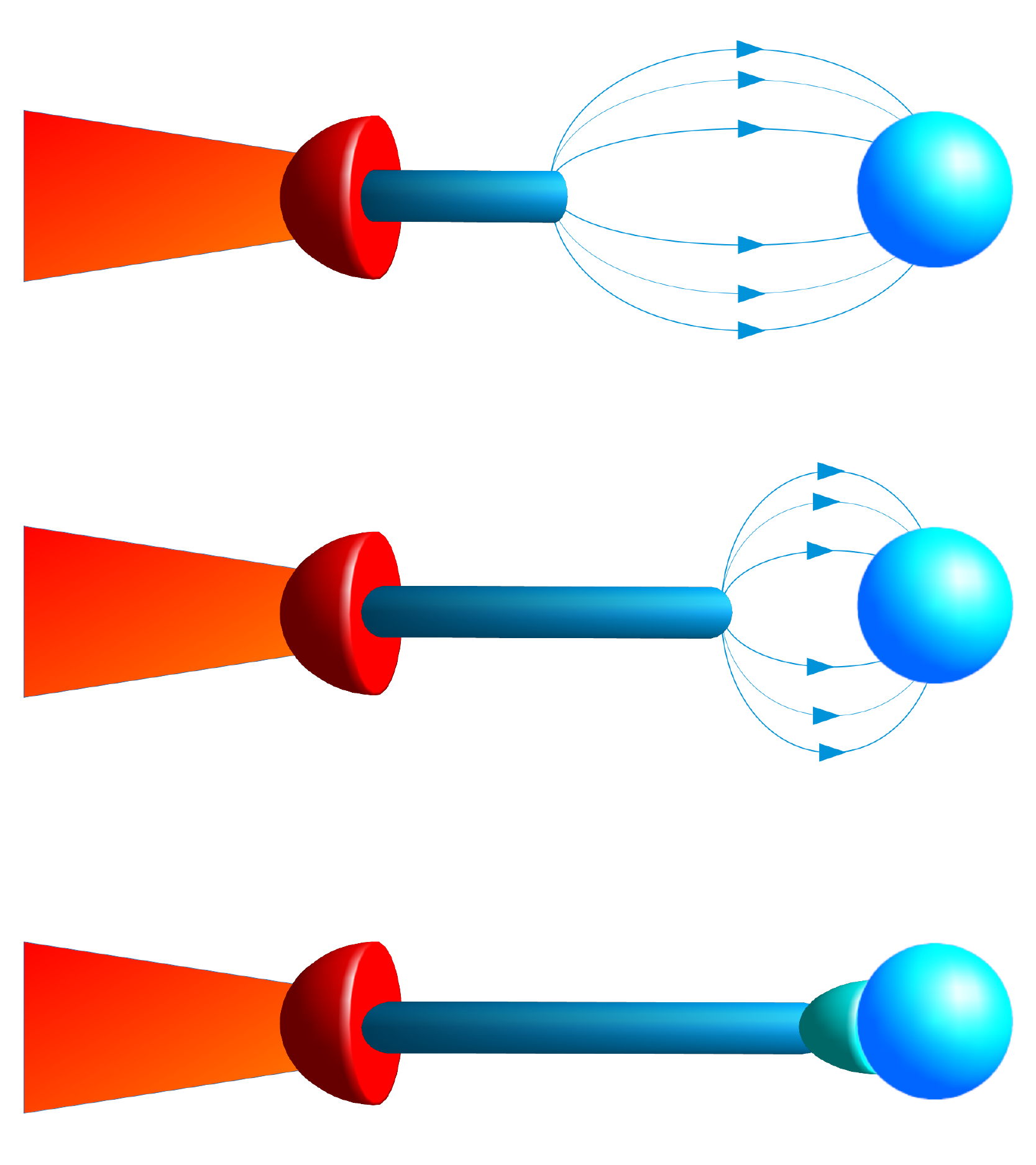}
\caption{{Cut in the plane of the star perpendicular to the left side of an e}dge-on accreting disk where the dust (red) destruction radius is further from the star than the gas disk (blue). Three possible scenarios are shown corresponding to decreasing strengths of the stellar magnetic field from top to bottom. Gas is channeled through the field lines according to MA (top and middle panels, corresponding to decreasing sizes of the magnetosphere), and directly onto the star through a BL (indicated in cyan at the bottom panel) in the absence of a strong enough magnetic field. }
\label{Fig:sketchMABL}
\end{figure}

{\subsection{Accretion in T Tauri Stars}}
\label{sect:TTs}
The initial approaches to understand accretion in CTTs assumed that the material falls directly from the disk to the star through a hot BL (Figure~\ref{Fig:sketchMABL}, bottom) where the angular momentum is drastically reduced and energy is released~\citep{Lynden74}. This BL perspective was successful in explaining the observed excesses both in the near-UV continuum and in the absorption lines (veiling) of CTTs, providing accretion estimates for these sources during the 1980s and 1990s~\citep{Bertout88,Hartigan91,Popham93,Hartigan95}. It was during this last decade and the beginning of the new century when the accretion paradigm changed into a magnetically driven perspective (see e.g.,~\citep{Valenti93}). According to the MA scenario (Figure~\ref{Fig:sketchMABL}, top and middle), the stellar magnetic field truncates the inner disk and channels the material at roughly free fall velocities until it shocks onto the star generating hot accretion spots that cover a few percent of the stellar surface~\citep{Uchida85,Konigl91,Shu94}. In fact, the $\sim$ kG magnetic fields observed in TTs were found to be strong enough to truncate the inner disk at a few stellar radii (see~\citep{Krull99,Krull02,Bouvier07} and Section~\ref{sect:MA_validity}), and the flux excesses could also be explained from the energy released in the accretion shocks~\citep{Gullbring98,Calvet98}. In addition, the MA scenario can address other phenomena that can be hardly interpreted from the BL view. First, although emission line profiles broadened by several hundred km s$^{-1}$ and with blueshifted self-absorptions or P Cygni profiles could in principle be explained from hot gas in Keplerian motion very close to the star\footnote{Please note that no BL model of line emission is yet available and it is not clear what exactly to expect from this scenario.} and outflowing material, CTTs also show redshifted self-absorptions and inverse P Cygni line profiles. These can only be explained from the presence of infalling gas in front of the central star, which can be observed in relatively inclined disks if accretion occurs at high latitudes under the MA geometry~\citep{Calvet92,Edwards94}. The redshifted self-absorptions and inverse P Cygni profiles of CTTs could indeed be reproduced from models assuming magnetically channeled accretion~\citep{Hartmann94,Muzerolle01}. Secondly, multi-epoch campaigns devoted to TT stars revealed periodicity that can be interpreted from stellar rotation and hot spots at the stellar surface generated in the MA shocks~\citep{Bertout88,Bouvier95}. Moreover, accreting CTTs seem to rotate slower than non-accreting WTTs, which can be indirectly explained from MA if the stars are initially locked to the inner disk Keplerian rotation through the magnetic channels~\citep{Edwards93,Bouvier95}. Although some controversy still remains concerning the ``disk-locking'' view (see e.g., the related discussion in~\citep{Landin16}), the main lines of evidence summarized above are supported by many independent works that reached a consensus about the validity of MA against BL. {A similar discussion about MA in CTTs as observed at short UV and X-ray wavelengths can be found in the review by Schneider et al. (2020)~\citep{Schneider20} for this same special issue of the journal.}   

Presently accretion rates in CTTs are commonly derived assuming MA, either from emission line or accretion shock modeling (e.g.,~\citep{Kurosawa06,Lima10,Ingleby13,Manara16})\footnote{It is noted that accretion rates based on the ``slab'' models introduced in Valenti et al. (1993)~\citep{Valenti93} can be interpreted both from the BL and MA scenarios, and are not the same as the MA shock models based on the work by Calvet \& Gullbring (1998)~\citep{Calvet98}. However, both provide roughly equivalent values of $\dot{M}_{\rm acc}$ for CTTs based on the Balmer continuum (e.g.,~\citep{Gullbring98,Herczeg08}).}, spectroscopic line veiling (e.g.,~\citep{Rei18} and references therein), or from the empirical correlations with the luminosity and width of emission lines (e.g.,~\citep{Natta04,Herczeg08,Fang09,Rigliaco12,Alcala14}). A~typical accretion rate for $\sim$1~M$_{\odot}$ TT stars is $\sim$10$^{-8}$~M$_{\odot}$ yr$^{-1}$, although $\dot{M}_{\rm acc}$ tends to increase with the stellar mass with an intrinsic spread of orders of magnitude.               

More specific historical approaches on accretion disks in TT stars and on how our understanding of accretion in these sources has evolved over time can be found in the literature (e.g.,~\citep{Bertout07,Basri07}), as well as detailed reviews on MA mainly devoted to low-mass stars~\citep{Bouvier07,Hartmann16}. 

{\subsection{Accretion in Herbig Ae/Be Stars}}
\label{sect:HAeBes}
Concerning the HAeBes, the initial evidence during the 1990s indicating that there is ongoing accretion was mainly based on UV emission lines or infrared excesses, although they provided somewhat contradictory results (see the review in~\citep{Perez94}). In particular, the accretion rates estimated from the IR excesses and the BL scenario by Hillebrand et al. (1992)~\citep{Hillenbrand92} were in the range 10$^{-5}$--10$^{-7}$ M$_{\odot}$ yr$^{-1}$, which are too high compared to more recent estimates (see below, Section~\ref{sect:BL}, and~\citep{Blondel06}). Sorelli et al. (1996)~\citep{Sorelli96} then suggested that the redshifted absorptions observed in the NaI {doublet} lines of some HAes showing {UXOr-like photo-polarimetric variability} could be explained in the context of MA. {A} few years later, the first indications that MA could be a valid scenario for the HAes as a whole---but not for HBes---were based on spectropolarimetry. In fact, Vink et al. (2002)~\citep{Vink02} found a difference between the H$\alpha$ spectropolarimetric signatures of HAe and HBe stars, which they interpreted as a transition from magnetically driven to direct, disk-to-star accretion depending on the spectral type. Similar spectropolarimetric studies including more stars and spectral lines have confirmed this change of behavior, which has been related to different accretion modes in CTTs and HAe stars on the one hand and in HBes on the other~\citep{Vink03,Mottram07,Ababakr17}. Almost in parallel, Eisner et al. (2004)~\citep{Eisner04} also suggested that there may be a transition from MA in late-type HAeBes to disk accretion in early-type sources, this time based on different inner disk geometries as inferred from near-infrared (IR) interferometry. That same year, Calvet et al. (2004)~\citep{Calvet04} presented results based on MA shock modeling applied to a small sample of ``intermediate-mass TT stars'' (IMTTs) with properties in-between TTs and HAeBes. They showed not only that the near-UV excesses and line profiles of these sources are consistent with the MA paradigm, but also that the empirical correlation between the accretion luminosities from MA shock modeling and the luminosity of the Br$\gamma$ emission line extends from CTTs to IMTTs. However, the first detailed MA line and shock modeling applied to reproduce the observations of a HAe star was in the seminal paper by Muzerolle et al. (2004)~\citep{Muzerolle04}. In this work several optical line profiles of the prototypical star UX Ori were reproduced from MA. Moreover, Muzerolle et al. (2004)~\citep{Muzerolle04} suggested that the near-UV excess flux observed in the Balmer region of the spectra ($\sim$3000--4500 \AA{}) of many HAeBes~\citep{Garrison78} could be explained from MA shock modeling in a similar way as for CTT stars, establishing a calibration relating the observed ``Balmer excess'' ($\Delta$D$_B${, see the top left panel of Figure~\ref{Fig:accmethods}}) in UX Ori with $\dot{M}_{\rm acc}$.

The previous works were then extrapolated to infer initial estimates of accretion rates based on MA for relatively wide samples of HAeBes. In particular, the empirical correlation with the Br$\gamma$ luminosity observed by Calvet et al. (2004)~\citep{Calvet04} was used by Garcia-Lopez et al. (2006)~\citep{GarciaLopez06} to infer $\dot{M}_{\rm acc}$ values for dozens of HAes. Although Calvet et al. (2004)~\citep{Calvet04} derived the correlation from a sample of IMTTs, a very similar correlation was later found for the HAeBes (see below), for which the extrapolation by Garcia-Lopez et al. (2006)~\citep{GarciaLopez06} {proved} accurate. Similarly, the $\Delta$D$_B$-$\dot{M}_{\rm acc}$ calibration by Muzerolle et al. (2004)~\citep{Muzerolle04} was also applied to estimate accretion rates of dozens of HAeBes with a wide range of stellar properties~\citep{Donehew11,Pogodin12}, although that calibration is only valid for stars with the same stellar parameters than UX Ori. As we will see next and in Section~\ref{sect:uv}, the extrapolation of such a calibration to HAeBe stars with different stellar properties can lead to accretion rates systematically biased by more than an order of magnitude.  

The first self-consistent estimates of $\dot{M}_{\rm acc}$ based on MA for a wide sample of HAeBes were made by Mendigut\'\i{}a et al. (2011)~\citep{Mendi11b} from the observed Balmer excesses of 38 northern stars. The photometric excess of each object was reproduced using the MA shock models of Calvet \& Gullbring (1998)~\citep{Calvet98}, deriving individual mass accretion rates for the whole sample and demonstrating that the calibration $\Delta$D$_B-\dot{M}_{\rm acc}$ is strongly dependent on the specific stellar properties. In particular, a~given $\Delta$D$_B$ translates into significantly higher accretion rates as the stars are hotter, and especially as the stellar surface gravity decreases (i.e., for smaller M$_*$/R$_*$ ratios). In addition, that paper showed that the widely used empirical calibrations between the accretion luminosity (L$_{\rm acc}$) and the luminosity of emission lines in CTTs can be extended to the HAeBes, at least for the H$\alpha$, [OI]6300, and Br$\gamma$ lines studied in that work (see also~\citep{Mendi13}). Interestingly, Mendigut\'\i{}a et al. (2011)~\citep{Mendi11b} also reported that it is impossible to reproduce the strong Balmer excesses of a few HBe stars in their sample from MA shock modeling, suggesting that an alternative accretion mechanism may {operate in} these objects and constituting a strong support to {the} initial claims from spectropolarimetry and interferometry mentioned above. Later, Fairlamb et al. (2015)~\citep{Fairlamb15} applied a similar methodology to derive accretion rates from MA shock modeling and X-Shooter spectra of 91 southern HAeBes. Again, {a significant fraction ($>$25$\%$)} of the HBe stars in that new sample could not be fitted from MA, reinforcing the view that the accretion physics could change for the stars with the earliest spectral types. The use of the X-Shooter spectra covering a wide wavelength range from the near-UV to the near-IR led Fairlamb et al. (2017)~\citep{Fairlamb17} to update previous empirical correlations from Mendigut\'\i{}a et al. (2011)~\citep{Mendi11b} and to find new ones between L$_{\rm acc}$ and the luminosity of many other emission lines, which are very similar to the corresponding correlations in CTTs~\citep{Alcala14}. The origin of these intriguing correlations in both CTTs and HAeBes was studied in Mendigut\'\i{}a et al. (2015)~\cite{Mendi15}. This work showed that indeed all lines from the near-UV to the near-IR can be used to infer accretion luminosities---even if some may not be physically related with accretion---because they reflect an underlying relation between L$_{\rm acc}$ and the stellar luminosity, L$_*$. In fact, the recent work by Wichittanakom et al. (2020)~\citep{Wichittanakom20} provides an empirical calibration between L$_{\rm acc}$ and L$_*$ for HAeBes that can also be used to {derive rough estimates of averaged} accretion rates. That the slope of the L$_{\rm acc}$-L$_*$ empirical calibration is shallower for HBes than for HAes and CTTs was interpreted by Wichittanakom et al. (2020)~\cite{Wichittanakom20} as the signature of a different physical mechanism driving accretion; MA in HAes, and BL in HBes. 

Currently, Gaia has allowed the characterization of hundreds of known HAeBe stars~\citep{Vioque18} and accretion rate estimates from the empirical correlations with the H$\alpha$ or the stellar luminosities are available for most of them~\citep{Arun19,Wichittanakom20,Guzman20}. The typical accretion rates of HAes from MA, $\sim$10$^{-7}$~M$_{\odot}$ yr$^{-1}$, are an order of magnitude larger than for CTTs, and there is again great scatter and a scaling relation with the stellar mass. 

It is noted that the fact that we can derive accretion rates of HAeBes based on MA does not mean that other paradigms cannot reproduce at least some of the observational properties of the same objects, and thus that they could also provide alternative accretion rates. Unfortunately, the BL view has received much less theoretical attention and only a few works provide accretion rate estimates of HAeBe stars based on this scenario~\citep{Hillenbrand92,Blondel06}. The following section discusses the main arguments supporting and challenging MA as a main driving mechanism in HAeBes, and in Sections~\ref{sect:BL} and~\ref{sect:uv} the BL measurements and how they compare to MA estimates are discussed.

\section{Is Magnetospheric Accretion Plausible for Herbig Ae/Be Stars?}
\label{sect:MA_validity}
So far we have seen that initial works suggested that the validity of MA could be extended at least to the late-type HAeBes, later supported by MA modeling that has led to direct measurements of accretion rates and to the extension of the empirical correlations as indirect probes of accretion. In addition, different lines of evidence support MA as the driving mechanism at least for HAe stars. For instance, multi-epoch analysis finds that the timescales and the extent of the H$\alpha$ variability is similar for objects ranging in mass from 0.1 to 5 M$_{\odot}$~\citep{Costigan14} but smaller for HBe stars~\citep{Mendi11a}. Statistical studies of wide samples of HAeBes in the optical and the near-IR show that the presence of redshifted and blueshifted self-absorptions in several emission lines are consistent with MA acting in HAes (although with small magnetospheres; see below) but not in HBes, for which the BL scenario seems more \mbox{suitable~\citep{Cauley14,Cauley15}}. Direct constraints of the Br$\gamma$ and H$\alpha$ line emitting regions from current spectro-interferometric facilities reveal that while those are in principle small enough to be consistent with the expected MA sizes in several HAeBes, many show more extended regions likely indicating an additional wind component (see e.g.,~\citep{Mendi17} and references therein). Analogous spectro-interferometric studies devoted to fainter T Tauri stars are less frequent (see e.g.,~\citep{Eisner14,Bouvier20}), but in fact their emission line profiles are currently reproduced using hybrid models that consider magnetically driven accretion and winds (e.g.,~\citep{Kurosawa06,Lima10}). Similarly, the emission line profiles of HAeBes can also be reproduced either from MA modeling alone or combined with magnetically driven winds~\citep{Muzerolle04,Mendi11b,Tambovtseva14,Tambovtseva16a,Tambovtseva16b,GarciaLopez16,Mendi17,Moura20}. Nonetheless, although many of these models are natural extensions from previous devoted to CTT stars, they lack a careful treatment of the {larger} projected rotational velocities that are characteristic of many HAeBes, which still constitutes a major limitation for the applicability of MA line modeling in these objects~\citep{Muzerolle01,Mendi17,Moura20}. Other direct and indirect observational tests mentioned in the previous section that were important to support MA in CTTs can hardly be applied to HAeBes. In particular, hot accretion spots cannot be generally observed because their temperatures and the stellar ones are comparable for HAeBes~\citep{Muzerolle04}. Moreover, the potential influence of accretion on the stellar rotation is also difficult to measure given that non-accreting ``weak HAeBes'' analogous to the WTTs are not {well identified} (but see Section~\ref{sect:MA} and~\citep{Hernandez05,Mendi11b,Alecian13b}).   

On the other hand, a few works suggest that MA could not be valid for HAeBes. For instance, the analysis of multi-epoch spectroscopic observations of two HAeBe stars suggest that the {absence of inverse P Cygni profiles} and {the stronger variability at longer timescales observed in the blueshifted absorptions of spectral lines} are inconsistent with a scaled-up T Tauri MA scenario~\citep{Aarnio17}. Similarly, that the fraction of redshifted and blueshifted absorption profiles in the HeI 10830 \AA{} line of 5 ``magnetic'' and 59 ``non-magnetic'' HAeBe stars is similar may indicate that the stellar magnetic field does not play a role in the gas kinematics~\citep{Reiter18}.   

In fact, the primary requirement for MA to {apply} is the presence of a strong enough magnetic field capable of truncating the inner disk and channeling the material towards the stellar surface. Magnetic fields of the order of $\sim$1 kG are commonly detected in CTT stars, based on $\geq$ a dozen of such sources for which measurements have been carried out (see~\citep{Bouvier07,Villebrun19} and references therein). Those strong magnetic fields are enough to sustain MA in CTTs (e.g.,~\citep{Krull99,Krull02,Bouvier07}). On the opposite, the vast majority of HAeBes show magnetic fields $\leq$ hundreds of G or below detection limits, as inferred from wide samples including dozens of sources (e.g.,~\citep{Donati97,Hubrig04,Wade07,Hubrig09,Alecian13,Jarvinen19a}). The contrast between both previous results is probably behind the strongest arguments against MA working in HAeBes, and it certainly motivates the question on whether this scenario is plausible for the HAeBe regime or not (see e.g., the corresponding discussions in~\citep{Wade07,Cauley14,Fairlamb15,Aguilar16,Moura20}).

Following Johns-Krull et al. (1999)~\citep{Krull99}, a lower limit of the magnetic field required to drive MA as a function of stellar and accretion parameters is given by:
\begin{equation}
\label{B_min}
\rm B_{min} \geq 1.1\times\left(\frac{M_*}{M_{\odot}}\right)^{(2/3)}\times\left(\frac{\it \dot{M}_{\rm acc}}{10^{-7}M_{\odot}yr^{-1} }\right)^{(23/40)}\times\left(\frac{R_*}{R_{\odot}}\right)^{-3}\times\left(\frac{P_*}{1 day}\right)^{(29/24)},
\end{equation}

(see~\citep{Cameron93}). The stellar rotation period P$_*$ can be expressed as 2$\pi$R$_{*}$$\sin$ {\it i}/v$\sin$ {\it i}, with v$\sin$ {\it i} the projected rotational velocity. Therefore, Equation~(\ref{B_min}) indicates that although the necessary B$_{min}$ increases with the stellar mass and accretion rate, larger for HAeBes than for CTTs, the major dependence is on the inverse of the stellar radius and rotational velocity. Given that both previous parameters tend to be substantially larger for more massive objects, the value of B$_{min}$ is typically smaller for HAeBes than for CTTs. In last term, this results from the fact that at distances below the co-rotation radius accretion dominates over the magnetic pressure driving winds, and such a radius tends to be comparatively smaller in HAeBes. 

Table~\ref{Table:Bfield} shows the compilation by Hubrig et al. (2015)~\citep{Hubrig15} listing a representative sample of HAeBes with {large-scale, organized magnetic fields averaged} from detections made by different teams and instrumentation (B$_{measured}$). The stellar and accretion parameters necessary to derive the minimum magnetic fields from Equation~(\ref{B_min}) are also shown. The comparison between the last two columns shows that B$_{measured}$ $\geq$ B$_{min}$ in most cases (considering the errorbars of the measured values), as expected if MA is the driving mechanism. There are only four possible exceptions: BF Ori, HD 101412, HD 104237, and HD 190073. However, the use of slightly different stellar parameters and accretion rates could translate into smaller values of B$_{min}$ and make them consistent with the B$_{measured}$ values. For instance, the stellar parameters and accretion rate for BF Ori in Mendigut\'\i{}a et al. (2011)~\citep{Mendi11b} lead to B$_{min}$ $\sim$50 G, smaller than the one measured and thus in potential agreement with MA too. Similarly, close binarity can introduce important uncertainties in the stellar parameters, accretion rates, and measured magnetic fields. In fact, HD 104237 is a spectroscopic binary with a HAe primary and a TT secondary and a recent work reports that the primary has a weak magnetic field ranging from $\sim$47 to 72 G~\citep{Jarvinen19b}. HD 101412 shows the largest difference between B$_{min}$ and B$_{measured}$, which is more than $\sim$ an order of magnitude smaller. Nonetheless, other measurements of the same star suggested that this could indeed be one of the HAeBes with the strongest magnetic field of several kG (see~\citep{Hubrig10} and references therein), which combined with uncertainties in stellar parameters and accretion rates could be potentially consistent with the minimum magnetic field necessary to drive MA. Finally, HD 190073 is a massive HBe star and the most recent measurements in J\"arvinen et al. (2019)~\citep{Jarvinen19a} confirm that its magnetic field is below $\sim$100~G, a factor $>$6 smaller than the value of B$_{min}$ estimated here and ruling out MA as the driving mechanism for this source.

Other analytical expressions making different assumptions of the coupling of the magnetic field with the inner disk yield constraints somewhat different than Equation~(\ref{B_min}) (see references in~\citep{Krull99,Krull02}). However, considering the uncertainties involved the typical stellar and accretion parameters of HAeBes lead to minimum magnetic fields that are comparable to the different measurements in the literature cited above; of the same order than the ones included in Table~\ref{Table:Bfield} for the representative sample of HAeBes with averaged detections from different teams. Concerning the HAeBes with non-detections, it must be emphasized that detection limits of several hundreds G are common~\citep{Wade07}. While such limits are not problematic for TTs with kG magnetic fields, they are of the same order than the magnetic fields necessary to drive MA in most HAeBes. Moreover, the uncertainties involved and the magnetic field detection limits are generally larger for HAeBes than for TTs. This is because, first, magnetic field estimates are based on measurements of different photospheric lines that are less abundant in HAeBes than in TTs, reducing the statistical reliability. Perhaps more importantly, rotational line broadening constitutes an important limitation in measuring the strength and configuration of stellar magnetic fields. In particular, Villebrun et al. (2019)~\citep{Villebrun19} recently showed that for v$\sin$ {\it i} $>$ 80 km s$^{-1}$ the detection limits could be large enough to prevent firm detections of large-scale magnetic fields, and such limits can be $>$1 kG for v$\sin$ {\it i} $>$ 50 km s$^{-1}$ for ordered, dipolar magnetic fields. In turn, HAeBes commonly show v$\sin$ {\it i} $\geq$ 100 km s$^{-1}$. Thus, it is unclear whether non-detections in HAeBes refer to an actual absence of a strong enough magnetic field to drive MA or not.

\begin{table}[H]
\caption{Magnetic fields of HAeBe stars.}
\label{Table:Bfield}
\centering
\begin{tabular}{lllllllll}
\toprule
\textbf{Star}	& \textbf{M}\boldmath{$_*^{min}$}	& \textbf{R}\boldmath{$_*^{max}$} & \textbf{v}\boldmath{$\sin$} \textbf{{\it \boldmath{i}}}\boldmath{$^{max}$} & \textbf{{\it i}}\boldmath{$^{min}$}& \textbf{P}\boldmath{$_*^{min}$} & \boldmath{$\dot{M}_{\rm acc}^{min}$} & \textbf{B}\boldmath{$_{min}$} &  \textbf{B}\boldmath{$_{measured}$}\\
\midrule
\dots& \textbf{M}\boldmath{$_{\odot}$}&\textbf{R}\boldmath{$_{\odot}$}& \textbf{km s}\boldmath{$^{-1}$}& \boldmath{$^{\circ}$}& \textbf{days}& \textbf{M}\boldmath{$_{\odot}$} \textbf{yr}\boldmath{$^{-1}$}& \textbf{G} & \textbf{G}\\
\midrule
HD 31648  	   & 1.9		  & 2.4 	      & 102		     & 39	   & 0.75 		    & 1.12~$\times$~10$^{-7}$  		       & 92  		       & 416~$\pm$~125\\
HD 35929           & 2.2		  & 7.6 	      & 64 		     & 32	   & 3.19 		    & 3.98~$\times$~10$^{-7}$  		       & 38  		       & 54~$\pm$~23 \\
HD 36112	   & 1.6		  & 2.2 	      & 59 		     & 49	   & 1.43 		    & 5.25~$\times$~10$^{-8}$  		       & 150 		       & 89~$\pm$~84 \\
V380 Ori           & 2.6		  & 4.0 	      & 7.8		     & 27	   & 4.20 		    & 3.02~$\times$~10$^{-7}$  		       & 347		       & 2120~$\pm$~150\\
BF Ori		   & 1.7		  & 2.0		      & 48 		     & 70	   & 1.98 		    & 5.75~$\times$~10$^{-8}$  		       & 326 		       & 87~$\pm$~36 \\
HD 58647	   & 3.4		  & 5.5 	      & 122		     & 55	   & 1.87 		    & 1.07~$\times$~10$^{-6}$  		       & 125 		       & 218~$\pm$~69 \\
Z CMa		   & 1.9		  & 11.8	      & 110		     & 30	   & 2.72 		    & 1.82~$\times$~10$^{-7}$  		       & 5 		       & 1231~$\pm$~164\\
HD 97048           & 2.3		  & 2.4 	      & 160		     & 38	   & 0.47 		    & 1.78~$\times$~10$^{-7}$  		       & 77  		       & 105~$\pm$~58 \\
HD 98922	   & 4.5		  & 13.6	      & 53 		     & 20	   & 4.44 		    & 2.75~$\times$~10$^{-6}$  		       & 49   		       & 135~$\pm$~64 \\
HD 100546	   & 2.2		  & 2.0		      & 55 		     & 22	   & 0.69 		    & 1.02~$\times$~10$^{-7}$  		       & 150 		       & 106~$\pm$~52 \\
HD 101412	   & 2.3		  & 2.8 	      & 4  		     & 80	   & 34.9		    & 2.29~$\times$~10$^{-7}$  		       & 10,291		       & 273~$\pm$~53 \\
HD 104237	   & 2.0	          & 3.0		      & 8  		     & 8 	   & 2.64 		    & 1.41~$\times$~10$^{-7}$  		       & 255 		       & 56~$\pm$~35 \\
HD 139614	   & $<$1.6		  & 1.4 	      & 27 		     & 32	   & 1.39 		    & 4.68~$\times$~10$^{-8}$  		       & $<$528		       & 73~$\pm$~26 \\
HD 144432	   & 1.6		  & 2.4 	      & 83 		     & 24	   & 0.60 		    & 5.75~$\times$~10$^{-8}$  		       & 42  		       & 100~$\pm$~50 \\
HD 144668          & 2.1		  & 3.9 	      & 210		     & 52	   & 0.74 		    & 3.24~$\times$~10$^{-7}$  		       & 42 		       & 106~$\pm$~34 \\
HD 150193	   & 2.0		  & 2.4 	      & 113		     & 32	   & 0.57 		    & 1.23~$\times$~10$^{-7}$  		       & 72  		       & 159~$\pm$~136\\
HD 176386	   & 2.4		  & 2.4 	      & 181		     & 50	   & 0.51 		    & 1.70~$\times$~10$^{-7}$  		       & 87  		       & 130~$\pm$~81 \\
HD 190073	   & 3.8		  & 12.5	      & 8  		     & 31	   & 40.8		    & 1.82~$\times$~10$^{-6}$  		       & 642		       & 62~$\pm$~21 \\

\bottomrule
\end{tabular}\\
\begin{tabular}{@{}c@{}} 
\multicolumn{1}{p{\textwidth -.88in}}{\footnotesize 
\textbf{Notes.} The listed minimum/maximum values for the stellar mass, radius, projected rotational velocity, disk inclination, stellar rotation period, and accretion rate (Cols. 2--7) are used to derive the minimum possible values of the magnetic field necessary to drive MA from Equation~(\ref{B_min}) (Col. 8). For comparison, the last column lists the {value of the large-scale, organized magnetic field} based on data from different teams and instrumentation~\citep{Hubrig15}. \textbf{References.} M$_*^{min}$, R$_*^{max}$, and $\dot{M}_{\rm acc}^{min}$ are taken from Guzm\'an-D\'\i{}az et al. (2020)~\citep{Guzman20}. v$\sin$ {\it i}$^{max}$ and {\it i}$^{min}$ are taken from Reiter et al. (2018) and references therein~\citep{Reiter18}, except for BF Ori~\citep{Mendi11b}; HD 58647~\citep{Mora01,Kurosawa16}; Z CMa~\citep{Hartmann09,Alonsoalbi09}; HD 97048~\citep{Wade07,Ginski16}; HD 98922 (inclination from~\citep{Aarnio17}); HD 100546~\citep{Acke04,Tatulli11} and HD 104237~\citep{Garcia13,Cowley13}. For HD 176386 an inclination of 50$^{\circ}$ is assumed. Rotational periods are derived from the values for R$_*^{max}$, v$\sin$ {\it i}$^{max}$ and {\it i}$^{min}$, except for V380 Ori~\citep{Alecian09}.}
\end{tabular}
\end{table}

In summary, although the absence of kG magnetic fields in HAeBes is commonly argued as a main flaw against MA working in these stars, such strong magnetic fields are necessary for lower-mass CTT stars but not for HAeBes, which require much smaller values to drive accretion magnetically. In fact, the minimum magnetic fields necessary to drive MA in HAeBes are generally consistent with current measurements and actual detection limits. In this respect, MA would remain as a plausible scenario also for HAeBes. Nonetheless, more work is still necessary concerning magnetic fields in HAeBes, in particular aiming at obtaining high signal-to-noise observations that reduce the detection limits and are capable of clarifying the somewhat contradictory results sometimes obtained by different teams and observational techniques (see e.g.,~\citep{Bagnulo12}). The magnetic field issue actually remains as a fundamental observational and theoretical problem that affects our understanding of how intermediate-mass stars grow. In turn, if MA is the driving mechanism in HAeBes their typically smaller magnetic fields and larger rotational velocities naturally lead to more compact magnetospheres than for CTTs. The typical disk truncation radius for HAeBes would be $\sim$2.5~R$_*$, either based on analytical formula depending on the magnetic field (e.g.,~\citep{Konigl91}) or on the co-rotation radius (that sets a maximum radial distance from which gas can be magnetically driven onto the star; e.g.,~\citep{Shu94}). That radius is $\sim$ two times smaller than the one typically estimated for CTT stars (e.g.,~\citep{Calvet98}). Eventually, the disk truncation radius could reach the stellar surface for small enough magnetic fields, when we would enter the BL regime (Figure~\ref{Fig:sketchMABL}).

\section{Magnetospheric Accretion Measurements of HAeBe Stars}
\label{sect:MA}
Although there is not yet a consensus about the physical validity of MA in HAeBe stars, this scenario is the one that has been more frequently used to derive accretion rates in these sources (but see Section~\ref{sect:BL}). In this section, the MA-based approaches from which accretion rates can (or cannot) be derived for HAeBes and the corresponding accuracies are discussed (see also Section~5 in the review of~\citep{Aguilar16} for a discussion on the lower limits).

\begin{figure}[H]
\centering
\includegraphics[width=15 cm]{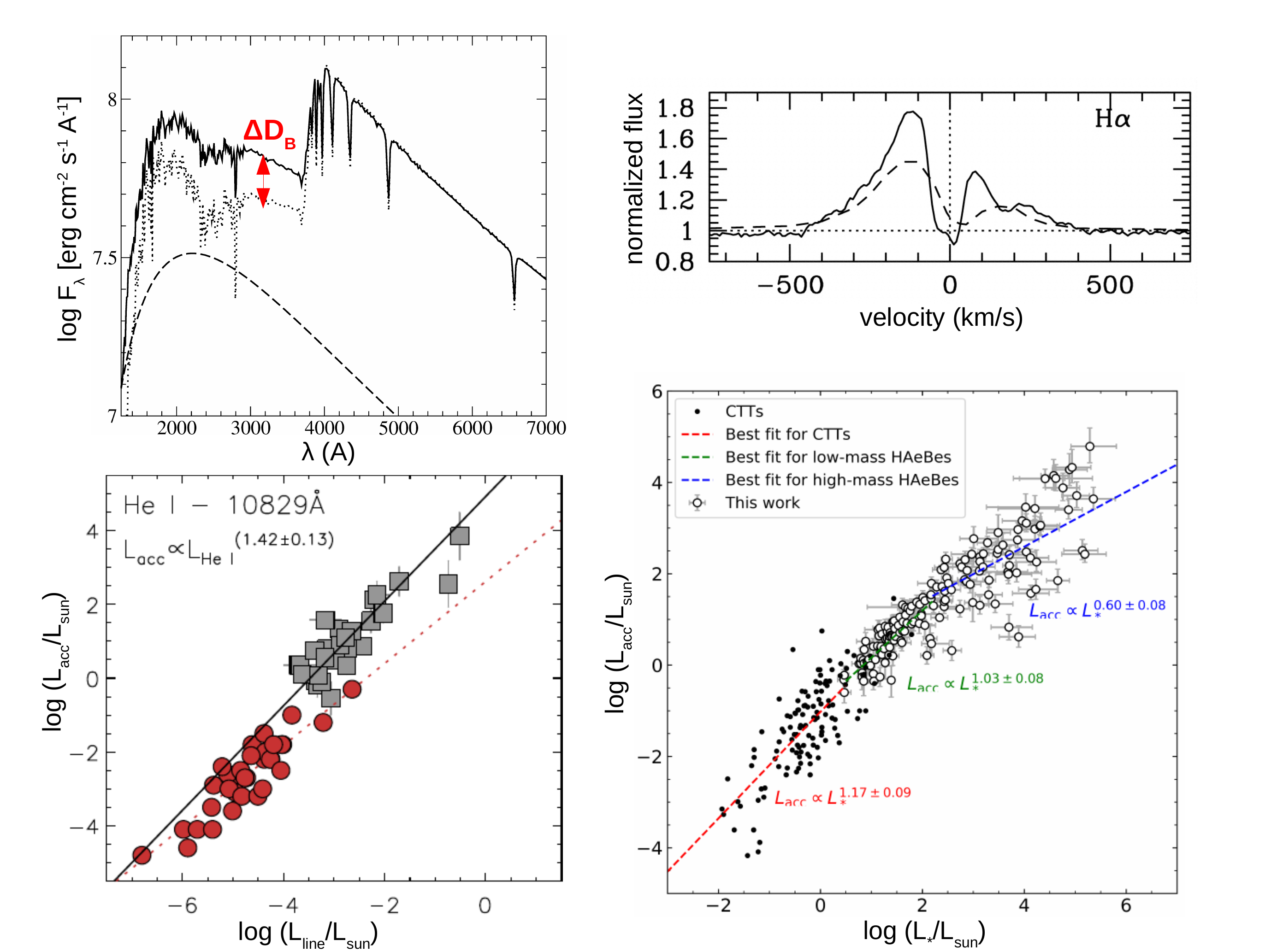}
\caption{(Adapted from several references, as indicated next). Figure exemplifying the different methods to estimate accretion rates of HAeBe stars based on the MA scenario. (\textbf{Top left}, from Mendigut\'\i{}a et al. (2011)~\citep{Mendi11b}, reproduced with permission © ESO.) The observed flux of a typical HAe star (solid line) shows an excess in the Balmer region of the spectrum ($\Delta$D$_B$) compared to the corresponding flux of a naked photosphere (dotted line), which can be modeled from the sum of such photospheric flux plus the contribution of the accretion shock (dashed line). (\textbf{Top right}, from Muzerolle et al. (2004)~\citep{Muzerolle04}, © AAS. Reproduced with permission.) Observed line profile of the HAe star UX Ori (solid line) versus the result of MA line modeling (dashed line) dependent on several stellar, disk, and geometrical parameters. (\textbf{Bottom left}, from Fairlamb et al. (2017)~\citep{Fairlamb17}) accretion luminosities of HAeBes (grey) versus the luminosity of the He I $\lambda$10829 \AA{} emission line. CTT measurements from Alcalá et al. (2014)~\citep{Alcala14} are overplotted in red. The corresponding best fits are indicated with the solid and dashed lines. (\textbf{Bottom right}, from Wichittanakom et al. (2020)~\citep{Wichittanakom20}) accretion luminosities of HAeBes (open symbols) versus stellar luminosities. CTT measurements from different references indicated in the original paper are overplotted with solid symbols. The best fits for the different regimes; CTTs, HAes, and HBes, are shown as indicated in the legend.}
\label{Fig:accmethods}
\end{figure}

As introduced in Section~\ref{sect:perspective}, accretion rates in HAeBes can be {observationally} estimated either from direct methods involving accretion shock or emission line modeling, or from indirect methods comprising empirical correlations with the previous{, model-based} accretion rates. Both the direct and indirect methods are exemplified in Figure~\ref{Fig:accmethods}. The direct methods are summarized in the~following.

\begin{itemize}
 \item Accretion shock modeling (Figure~\ref{Fig:accmethods}, top left): This is the most accurate method to directly infer a~value of $\dot{M}_{\rm acc}$ by reproducing the observed excess in the near-UV region of the spectrum from the contribution of the accretion shock. Such a contribution can be modeled as a blackbody for HAeBes~\citep{Muzerolle04,Mendi11b}, and its influence depends on free parameters {relatively constrained by theory, like} the inward flux of energy carried by the accretion funnels and the fraction of the stellar surface covered by the accretion shocks. The wavelength region of study in HAeBes has so far been centered on the Balmer jump, $\Delta$D$_B$, but it could be extended to shorter wavelengths (Section~\ref{sect:uv}, and see~\citep{Ingleby13} for TTs). The value of $\Delta$D$_B$, and thus the inferred value of $\dot{M}_{\rm acc}$, is itself distance-independent, but it could indirectly depend on the distance given that the measured excess depends on the stellar parameters assumed. Considering usual uncertainties and dependencies, the typical errorbar for $\dot{M}_{\rm acc}$ as estimated from this method is $<$0.5 dex. Details about the procedure, uncertainties involved, and measurements for dozens of northern and southern HAeBes can be consulted in the literature~\citep{Mendi11b,Fairlamb15}.
 \item Emission line modeling (Figure~\ref{Fig:accmethods}, top right): According to MA several emission lines like the Balmer series, the sodium doublet, or HeI transitions are at least partially generated in the hot gas free-falling within the magnetic channels that connect the inner disk and the accretion shocks on the stellar surface. Radiative transfer is applied normally assuming a simple dipole geometry for the magnetic field. Apart from the stellar parameters (T$_*$, M$_*$, R$_*$, and rotational velocity), MA line modeling depends on the disk inclination, the size of the magnetosphere---i.e., the disk truncation radius---and the gas temperature, on top of $\dot{M}_{\rm acc}$ which is normally the parameter that one wants to determine. Even in the best case scenario when the stellar parameters and geometry are well constrained, so far we can only derive upper limits on the disk truncation radius based on spectro-interferometry, although more direct constraints can be inferred from spectro-polarimetry for a few stars (the potential of this latter technique to probe such small scales is discussed in~\citep{Vink05a,Vink05b}). In turn, rough estimates of the gas temperature are solely based on empirical constraints but theory is still lacking in this respect~\citep{Muzerolle01}. Given the number of free parameters, MA line modeling normally serves to estimate $\dot{M}_{\rm acc}$ within $\sim$ an order of magnitude accuracy, although for well-known sources and a careful modeling the uncertainty can be significantly reduced (see e.g., the recent work for a TT star in~\citep{Thanathibodee20}). As noted in Section~\ref{sect:MA_validity}, MA line modeling including a complete treatment of the high rotational velocities that are typical in many HAeBes is still pending. Examples and details of line modeling applied to HAeBes, either considering MA alone or combined with magnetically driven winds, can be found in the literature~\citep{Muzerolle04,Mendi11b,Tambovtseva14,Tambovtseva16a,Tambovtseva16b,GarciaLopez16,Mendi17,Moura20}.
\end{itemize}

The indirect methods are discussed next:

\begin{itemize}
 \item Empirical correlations with the luminosity of emission lines (Figure~\ref{Fig:accmethods}, bottom left): The accretion luminosities of HAeBes inferred from the direct methods described above (L$_{\rm acc}$ $\sim$ GM$_*$$\dot{M}_{\rm acc}$/R$_*$) correlate with the luminosities of dozens of emission lines spreading from the near-UV to the near-IR, as previously found for CTTs (the extension of such correlations to specific lines at shorter and longer wavelengths in the far-UV and the mid-IR can be found e.g., in~\citep{Yang12,Rigliaco15} respectively). Regardless of the physical origin of those lines all can be used to derive accretion rates by measuring the emission line luminosity and using the corresponding empirical expression with the form log (L$_{\rm acc}$/L$_{\odot}$) = A(~$\pm$~A) + B(~$\pm$~B) ~$\times$~ log (L$_{line}$/L$_{\odot}$)~\citep{Mendi15}. The errors for the slopes and intercepts are determined from the least-square fits to the L$_{\rm acc}$-L$_{line}$ data, and the final typical uncertainty for $\dot{M}_{\rm acc}$ (once transformed from L$_{\rm acc}$ using the above mentioned formula and the stellar parameters) is $\sim$ 1 dex. This uncertainty could in principle be reduced by averaging the results obtained from different emission lines~\citep{Rigliaco12}. The most recent L$_{\rm acc}$-L$_{line}$ empirical expressions for HAeBes including dozens of emission lines can be found in Fairlamb et al. (2017)~\citep{Fairlamb17}, and accretion rates inferred from this method (in particular, from the correlation with L$_{H\alpha}$) can also be found in the literature for hundreds of HAeBes~\citep{Arun19,Wichittanakom20}.
 \item Empirical correlations with the stellar luminosity (Figure~\ref{Fig:accmethods}, bottom right): As mentioned above the fact that the luminosity of a given emission line correlates with L$_{\rm acc}$ does not necessarily mean that there is an actual physical link between the origin of the line and the accretion process, and such a correlation naturally results from the underlying one between L$_{\rm acc}$ and L$_{*}$~\citep{Mendi15}. Because the scatter in the latter correlation is similar than for the L$_{\rm acc}$-L$_{line}$ correlations, $\sim$ ~$\pm$~1 dex, accretion rates can be similarly derived from the L$_{\rm acc}$-L$_{*}$ empirical expression. This expression depends on the mass regime as described in Wichittanakom et al. (2020)~\citep{Wichittanakom20} including HAeBes, and has been recently used to derive accretion rates for almost all HAeBes known~\citep{Guzman20}.    
\end{itemize}

Concerning the indirect methods, these do not generally serve to probe accretion variability except strong changes over relatively long timescales. The intrinsic scatter of $\sim$ ~$\pm$1 dex that affects the L$_{\rm acc}$-L$_{line}$ correlations {limits} the range of accretion rate variations that could be analyzed. {Still, L$_{line}$ variations homogeneously measured for a given star could be potentially linked to accretion rate changes. However}, emission lines can be originated from other processes apart from accretion and thus their variability traces additional mechanisms and timescales. A good example is the bulk of the [OI]6300 emission line, which in HAeBes mainly probes to the low-density, surface layers of the protoplanetary disks~\citep{Acke05} although its luminosity still correlates with L$_{\rm acc}$ and serves to provide a rough estimate of the accretion rate~\citep{Mendi11b}. In fact, the simultaneous monitoring of the near-UV excess---directly related to accretion---and several line luminosities show that they can evolve differently {in HAeBes} (e.g.,~\citep{Mendi11b,Mendi13}). The careful monitoring of specific {CTTs} actually reveals significant time delays between the moment when the material shocks onto the stellar surface and the time when such an accretion event is reflected by several emission lines at different wavelengths~\citep{Dupree12}. 

On the other hand, the previously discussed direct and indirect methods to estimate accretion rates in HAeBes can be considered an extension from similar methodologies previously applied to CTT stars. However, it is worth mentioning that not all methods used in the low-mass regime can be extended to the HAeBes, which is summarized as follows:

\begin{itemize}
 \item Spectroscopic line veiling: CTTs show optical photospheric absorption lines smaller in depth than observed in WTTs or low-mass stars in the MS, which can be explained from the contribution of the hot accretion shock. Indeed, by removing the contribution of the stellar photosphere to the observed photospheric lines one can directly infer a value of $\dot{M}_{\rm acc}$ (see e.g.,~\citep{Hartigan89,Hartmann90,White03,White04} and references therein). In contrast, optical spectroscopic line veiling is not commonly observed in HAeBes, not because they are not accreting but due to the fact that the temperature of the accretion shocks is comparable or smaller to that of the stellar photosphere ($\sim$10000 K) and therefore the contrast effect is negligible (see~\citep{Muzerolle04}  for more details). Thus, optical spectroscopic line veiling is not a method to routinely derive accretion rates of HAeBes excepts perhaps for the coldest sources (see an example in~\citep{Mendi14}).   
 \item H$\alpha$ line width: The H$\alpha$ line width at 10$\%$ of peak emission, W$_{10}$ (H$_{\alpha}$), not only serves as a~qualitative indicator of accretion in CTTs and brown dwarfs~\citep{White03} but also as a rough quantitative estimator through an empirical correlation with $\dot{M}_{\rm acc}$~\citep{Natta04}. Based on the study of a HBe source, Boley et al. (2009)~\citep{Boley09} already suggested that such an empirical correlation may not extend to higher masses, which was later confirmed by Mendigut\'\i{}a et al. (2011)~\cite{Mendi11b} from the study of a larger sample of HAeBe stars. This work showed that while their typically large v$\sin$ {\it i} values are reflected by H$\alpha$ emission broadening in possible agreement with MA, the influence of rotation is that important that the $\dot{M}_{\rm acc}$-W$_{10}$ (H$_{\alpha}$) correlation breaks and thus cannot be used for the HAeBe~regime.   
\end{itemize}

In summary, most---but not all---methods that are commonly used to derive accretion rates in the low-mass regime and are based on MA can be extended to the HAeBes with similar accuracies. 

\section{Boundary Layer Measurements of HAeBe Stars}
\label{sect:BL}
As we have seen almost all efforts devoted to derive accretion rates in HAeBe stars rely on the MA paradigm, although the first estimates were actually based on the BL scenario and the observed near-IR excesses (see Section~\ref{sect:perspective} and \citep{Hillenbrand92}). However, such a wavelength range does not reflect accretion in most sources (see e.g., the discussions in~\citep{Muzerolle04,Mendi12} and references therein). Alternatively, Blondel \& Tjin A Djie (2006)~\citep{Blondel06} derived accretion rates of a relatively wide sample of HAeBes based on the observed UV excesses and the BL scenario. In this work, low resolution spectra from the International Ultraviolet Explorer (IUE) combined with optical information of dozens of late-type HAeBes (HAes and IMTTs) were reproduced in terms of the photosphere of a central star, an optically thick accretion disk, and a hot BL in-between the previous. As a result of such a modeling, stellar parameters and accretion rates were derived for most of the stars in that sample, constituting a unique database that can be compared to MA estimates mentioned above. 

A direct comparison between the accretion rates derived by Blondel \& Tjin A Djie (2006)~\citep{Blondel06} from BL and the ones inferred from MA shock modeling by Mendigut\'\i{}a et al. (2011) and Fairlamb et al. (2015)~\citep{Mendi11b,Fairlamb15} suggests that MA estimates tend to be smaller than from BL for many stars, in agreement with a similar comparison carried out for TTs~\citep{Gullbring98}. However, the stellar parameters and distances used in the BL and MA works differ---sometimes significantly---which could affect the previous conclusion. An alternative approach to compare accretion rates from MA and BL, avoiding the influence of stellar parameters and distances, is discussed next. 

\begin{figure}[H]
\centering
\includegraphics[width=13 cm]{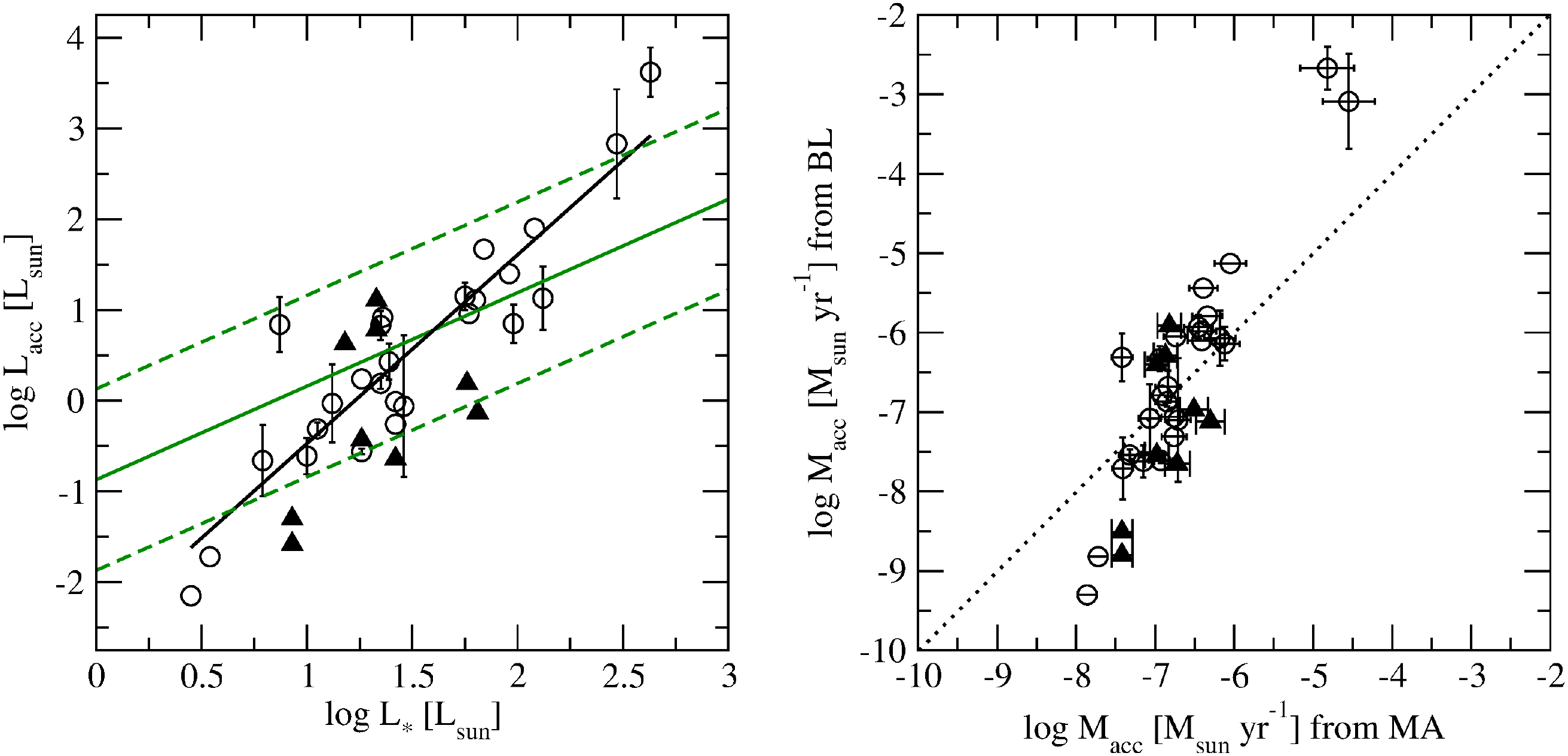}
\caption{(\textbf{Left panel}) Accretion luminosity versus stellar luminosity as derived in Blondel \& Tjin A Djie (2006)~\cite{Blondel06} from BL for a sample of HAe stars. The best fit is indicated with the black line (slope $\sim$ 2). The trend ($\pm$1~dex) that HAes (and TTs) follow when accretion luminosities are estimated from MA is indicated with the green lines (slope $\sim$ 1). (\textbf{Right panel}) Comparison between mass accretion rates as inferred from BL and MA for the same sample. Equal values are indicated with the dotted line. (\textbf{Both panels}) The black triangles represent lower limits for the accretion estimates from BL. }
\label{Fig:BL_vs_MA}
\end{figure}

Figure~\ref{Fig:BL_vs_MA} (left) shows the accretion luminosities inferred from the values of $\dot{M}_{\rm acc}$, M$_*$ and R$_*$ tabulated in Blondel \& Tjin A Djie (2006) (see Table 2 in~\citep{Blondel06}) versus the L$_*$ values derived in that same work. In the BL formalism, L$_{\rm acc}$ $\sim$ (1/2) ~$\times$~ GM$_*$$\dot{M}_{\rm acc}$/R$_*$ (see Section~\ref{sect:uv}). The $\dot{M}_{\rm acc}$ values {used in this formula} are the ones tabulated {in Blondel \& Tjin A Djie (2006)~\citep{Blondel06}} with inclinations to the line of sight different than 90$^{\circ}$, when available{. The edge-on values from that paper are taken} otherwise, which can be considered to be lower limits (triangles in Figure~\ref{Fig:BL_vs_MA}). In addition, for the stars with different estimates the averages are computed, the errorbars indicating the largest difference with respect to the individual values. The fit to the L$_{\rm acc}$-L$_*$ values of HAes when accretion is estimated from the BL view in Blondel \& Tjin A Djie (2006)~\citep{Blondel06} is indicated with a black line and has a slope $\sim$2. In contrast, the L$_{\rm acc}$-L$_*$ trend followed by HAes when their accretion luminosities are inferred from MA is overplotted in green. This is equal within errorbars to the one followed by TTs and has a smaller slope $\sim$1 (see the bottom right panel of Figure~\ref{Fig:accmethods} and~\citep{Mendi15,Wichittanakom20} for details).

Figure~\ref{Fig:BL_vs_MA} (right) compares the $\dot{M}_{\rm acc}$ values directly estimated by Blondel \& Tjin A Djie (2006)~\citep{Blondel06} from BL with the values inferred from the MA-based correlation between L$_{\rm acc}$ and L$_*$ (see Section~\ref{sect:MA}). In particular, the expressions from Wichittanakom et al. (2020)~\citep{Wichittanakom20} were used: log (L$_{\rm acc}$/L$_{\odot}$) = ($-$0.87 ~$\pm$~ 0.11) + (1.03 ~$\pm$~ 0.08) ~$\times$~ log (L$_*$/L$_{\odot}$) for L$_*$ $<$ 194 L$_{\odot}$, and log (L$_{\rm acc}$/L$_{\odot}$) = (0.19 ~$\pm$~ 0.27) + (0.60 ~$\pm$~ 0.08) ~$\times$~ log (L$_*$/L$_{\odot}$) for the two brightest stars. The original data for L$_*$ from Blondel \& Tjin A Djie (2006)~\citep{Blondel06} were used to infer the L$_{\rm acc}$ values, as well as their data for M$_*$ and R$_*$ to derive the final MA-based values for $\dot{M}_{\rm acc}$ = L$_{\rm acc}$R$_*$/GM$_*$. The comparison between BL- and MA-based estimates of $\dot{M}_{\rm acc}$ shows that the former tend to be larger for stars accreting at relatively high rates, which constitute $\sim$ 41$\%$ of the stars in the sample analyzed. On the opposite, the weak accretors show $\dot{M}_{\rm acc}$ (BL) $<$ $\dot{M}_{\rm acc}$ (MA), representing $\sim$ 21$\%$ of the stars in the sample. The rest of the stars, $\sim$ 31$\%$, show accretion rates that do not differ significantly regardless of the accretion scenario.

The mismatch between MA and BL estimates is in last term due to differences between the underlying assumptions concerning energy transformation under both approaches (e.g.,~\citep{Wichittanakom20}). In MA the gravitational energy is carried through the accretion funnels and released at hot spots in the stellar surface, whereas in BL the kinetic energy of the rotating disk heats the BL when material is drastically decelerated in this region. Further details about BL in comparison to MA are discussed in the next~section.

\section{The Ultraviolet Link}
\label{sect:uv}
The analysis in the previous section illustrates the importance of adopting a given accretion paradigm to interpret the data, as different scenarios lead to different values of the mass accretion rates. In turn, this has a potential effect on the shape of the L$_{\rm acc}$-L$_*$ correlation (or the roughly equivalent one between $\dot{M}_{\rm acc}$ and M$_*$) and other related scaling relations (Figure~\ref{Fig:BL_vs_MA} left;~\citep{Muzerolle03,Mendi15,Manara16b,Andrews18}, and references therein), the value of accretion-based disk masses~\citep{Hartmann98b,Andrews_Williams07,Mendi12,Dong18,Guzman20}, the disk dissipation timescales and driving processes including planet formation (e.g.,~\citep{Lin96,Alexander07,Najita07,Aguilar10,Mendi12,Manara18}, and references therein), the removal of angular momentum (e.g.,~\citep{Bodenheimer95,Stassun03}, and references therein), or even on the estimates of SFRs characterizing large star-forming regions when inferred from the sum of individual stellar accretion rates~\citep{Padoan14,Mendi18}, to cite some. Although so far the bulk of the evidence supports MA as the driving mechanism in HAe stars, the possibility that the alternative BL scenario plays a role in these objects, and especially in HBe stars, needs to be further explored. In this section, simple MA and BL shock models are compared to argue that the study of the UV wavelength region could contribute to discriminate between both competing accretion scenarios. 

The BL is geometrically described as an annulus around the star with radial thickness $\delta$. Following Lynden-Bell \& Pringle (1974)~\citep{Lynden74}
\begin{equation}
\label{Eq:delta}
\delta\propto\nu^{2/3}\left(\frac{G\rm M_*}{\rm R_{*}^{2}}\right)^{-1/3}\left(1-\left(\frac{\Omega_{*}}{\Omega_{K}}\right)^{2}\right)^{-1/3}= \delta_{0}{\rm R_*} \left(1-\left(\frac{\Omega_{*}}{\Omega_{K}}\right)^{2}\right)^{-1/3}. 
\end{equation}

The stellar and Keplerian angular velocities (at the star`s surface, i.e., the break-up velocity; $\Omega_{*}$/$\Omega_{K}$ $<$ 1) are  $\Omega_{*}$ = v$\sin$ {\it i}/R$_*$$\sin$ {\it i} and $\Omega_{K}$ = (GM$_*$/R$_{*}^{3}$)$^{1/2}$, respectively. The right term in Equation~(\ref{Eq:delta}) is derived assuming that the disk viscosity can be expressed as $\nu$ $\propto$ (GM$_{*}$R$_{*}$)$^{1/2}$~\citep{Lynden74,Tylenda77}, and it reflects the fact that the radial thickness of the BL increases for higher stellar to Keplerian velocity ratios. The exact value of $\delta_0$ (and therefore of $\delta$) is difficult to constrain observationally. Following Blondel \& Tjin A Djie (2006)~\citep{Blondel06}, it will be assumed a typical (median) value $\delta_0$ = 0.03 by default, although this is a free input parameter in principle ranging from 0.01 to 0.5 (see e.g.,~\citep{Popham93}).

Concerning the energy balance, {it is assumed that the kinetic energy of the disk material rotating Keplerian is transformed to heat the BL~\citep{Lynden74,Wichittanakom20}}, which {in zeroth order approach is represented by} an optically thick region characterized by a single blackbody temperature T$_{\rm BL}$ and emitting area 4$\pi$R$_{*}$$\delta$  
\begin{equation}
\label{Eq:TBL}
{\rm L_{acc}} = \left(\frac{1}{2}\right)\times\frac{{\rm GM_{*}}\dot{M}_{\rm acc}}{{\rm R_{*}}}=4\pi {\rm R_{*}}\delta \sigma {\rm T_{BL}^{4}},
\end{equation}

(see~\citep{NarayanPop94,Blondel06}). Please note that all the energy is assumed to be released in the BL, and the part devoted to spin-up the star is neglected. This is observationally justified considering that the majority of A and B stars rotate much slower than the break-up velocities at least at the beginning of the MS~\citep{Abt02,Zorec12}.

The total (non-extinct) flux per wavelength unit coming from the star-disk system ({\it F$_{\lambda}$}) can be divided into three components: the flux from the boundary layer ({\it F$_{\lambda}^{BL}$}), the photosphere of the star ({\it F$_{\lambda}^{*}$}) and the disk ({\it F$_{\lambda}^{disk}$})
\begin{equation}
\label{Eq:totflux}
F_{\lambda} = F_{\lambda}^{BL} + F_{\lambda}^{*} + F_{\lambda}^{disk}.
\end{equation}

Given that the contribution of the disk in the UV is negligible, it will not be considered here anymore. The first term in the previous equation can be expressed as 
\begin{equation}
\label{Eq:BL}
F_{\lambda}^{BL} = (\pi + 2\gamma)\cos i\frac{\delta \rm R_*}{d^2}B_{\lambda}({\rm T_{BL}}),
\end{equation}

(see~\citep{Tylenda77,Bertout88}), where {\it B}$_{\lambda}({\rm T_{BL}})$ is the blackbody radiance characterizing the boundary layer. The factor $(\pi + 2\gamma)$$\cos {\it i}$ describes the effect of partial screening of the boundary layer by the stellar surface due to inclination of the system in the line of sight, with $\gamma$ derived by geometrical arguments as
\begin{equation}
\label{Eq:gamma}
\sin \gamma = \left(\frac{1}{\sin i}\right)\sqrt{1-\left(\frac{\rm R_*}{{\rm R_*} + \delta /2}\right)^{2}},
\end{equation}

(see~\citep{Tylenda77}). Therefore, the BL flux received increases for larger radial thicknesses and for inclinations closer to pole-on, when the BL emitting region is less occulted by the stellar surface. Regarding the second term in Equation~(\ref{Eq:totflux}), this is
\begin{equation}
\label{Eq:star}
F_{\lambda}^{*} = F_{\lambda}^{phot}\left(\frac{\rm R_*}{d}\right)^2\left(1-\frac{\delta}{{\rm R_*}}\sin i\right),
\end{equation} 
where the photospheric flux is corrected by the fraction of the stellar surface that is covered by the BL annulus. This is supposed to have a physical height similar to its radial thickness $\delta$~\citep{Bertout88,Blondel06}. Therefore, the photospheric flux received decreases for inclinations closer to edge-on, when the stellar surface is more occulted by the BL.

For a given star with stellar parameters M$_{*}$, R$_{*}$ (or log g), projected rotational velocity v$\sin {\it i}$ and inclination to the line of sight {\it i}, the thickness of the boundary layer $\delta$ is estimated from Equation~(\ref{Eq:delta}). From this, different mass accretion rates provide different boundary layer temperatures through Equation~(\ref{Eq:TBL}). Then the flux contribution of the boundary layer and that of the stellar surface (using a Kurucz template with a given stellar temperature and surface gravity) is derived through Equations~(\ref{Eq:BL}) and~(\ref{Eq:star}), respectively. Finally, the expected total flux is obtained from Equation~(\ref{Eq:totflux}). As mentioned above, here we will focus on the UV range, for which the disk emission in the previous equation can be~neglected.

Concerning MA shock modeling, the same prescriptions described in Mendigut\'\i{}a et al. (2011)~\cite{Mendi11b} will be used here in order to compare the MA and BL predictions in the UV. There are two major differences with respect to the BL model described above. First, in the BL model the emitting region due to accretion is an equatorial annulus with thickness $\delta$ around the star, and the received flux depends on the inclination of the source. In contrast, in MA the emitting region is described from the accretion spots assumed to be located at high latitudes (i.e., inclination independent), and the corresponding fraction of the stellar surface covered by such spots (``filling factor''). Secondly, for a fixed value of $\dot{M}_{\rm acc}$ and a given set of stellar parameters, the free parameters in the MA model are the inward flux of energy carried by the accretion columns, ($\cal F$, normally expected to be around \mbox{10$^{12}$ erg cm$^{-2}$ s$^{-1}$}; see~\citep{Muzerolle04}) and the disk truncation radius (R$_i$, that depends on the specific star but it is typically $\sim$2.5~R$_*$ for HAeBes, see Section~\ref{sect:MA}). These parameters determine the temperature and filling factor characterizing the accretion columns and shocks (larger values of $\cal F$ and R$_i$ provide larger values of T$_{col}$ and smaller filling factors, respectively). In turn, in the BL model the free parameter is the size of the emitting region inferred from $\delta_0$, which also determines the BL temperature (larger values of $\delta_0$ provide smaller values of T$_{\rm BL}$ and vice versa). The interested reader can consult Mendigut\'\i{}a et al. (2011)~\cite{Mendi11b} for more details on the MA shock model.

Here, the MA and BL predictions in two regions of the UV spectra will be compared to each other. The excess in the Balmer region is defined as
\begin{equation}
\label{Eq:Balmerexcess}
\Delta D_{B} = 2.5 {\rm log} \left( \frac{F_U}{F_U^{phot}} \right),
\end{equation} 
where {\it F$_U$}/{\it F$_U^{phot}$} represents the ratio between the mean total, accretion-contributed flux and the mean photospheric flux in the wavelength region 3500--3700 \AA{} (i.e., the $U$ photometric band), where both are normalized at around 6000 \AA{} (V-photometric band). Kurucz templates are again used to represent $F_U^{phot}$. Analogously, to quantify the excess at a shorter wavelength it is defined
\begin{equation}
\label{Eq:UVexcess}
\Delta_{UV} = 2.5 {\rm log} \left( \frac{F_{2125}}{F_{2125}^{phot}} \right),
\end{equation}     
with {\it F$_{2125}$}/{\it F$_{2125}^{phot}$} being the ratio between the mean total, accretion-contributed flux and the mean photospheric flux in the wavelength region 2000--2250 \AA{}. Such a wavelength region has been chosen arbitrarily only for comparison purposes with $\Delta$D$_{B}$, but a different UV range could be analyzed instead. It should be noted that the effect of extinction is not considered either in the previous definitions or for the subsequent discussion based on analytical expressions. In addition, and assuming 10$\%$ and 5$\%$ errors in the total and photospheric fluxes, a typical uncertainty for the above defined excesses is $\sim$0.1 magnitudes, which represent the maximum errorbars that could be derived from broadband photometry. Such errors could be significantly reduced if e.g., high precision photometry or spectra are~used.

\begin{figure}[H]
\centering
 \includegraphics[width=12cm]{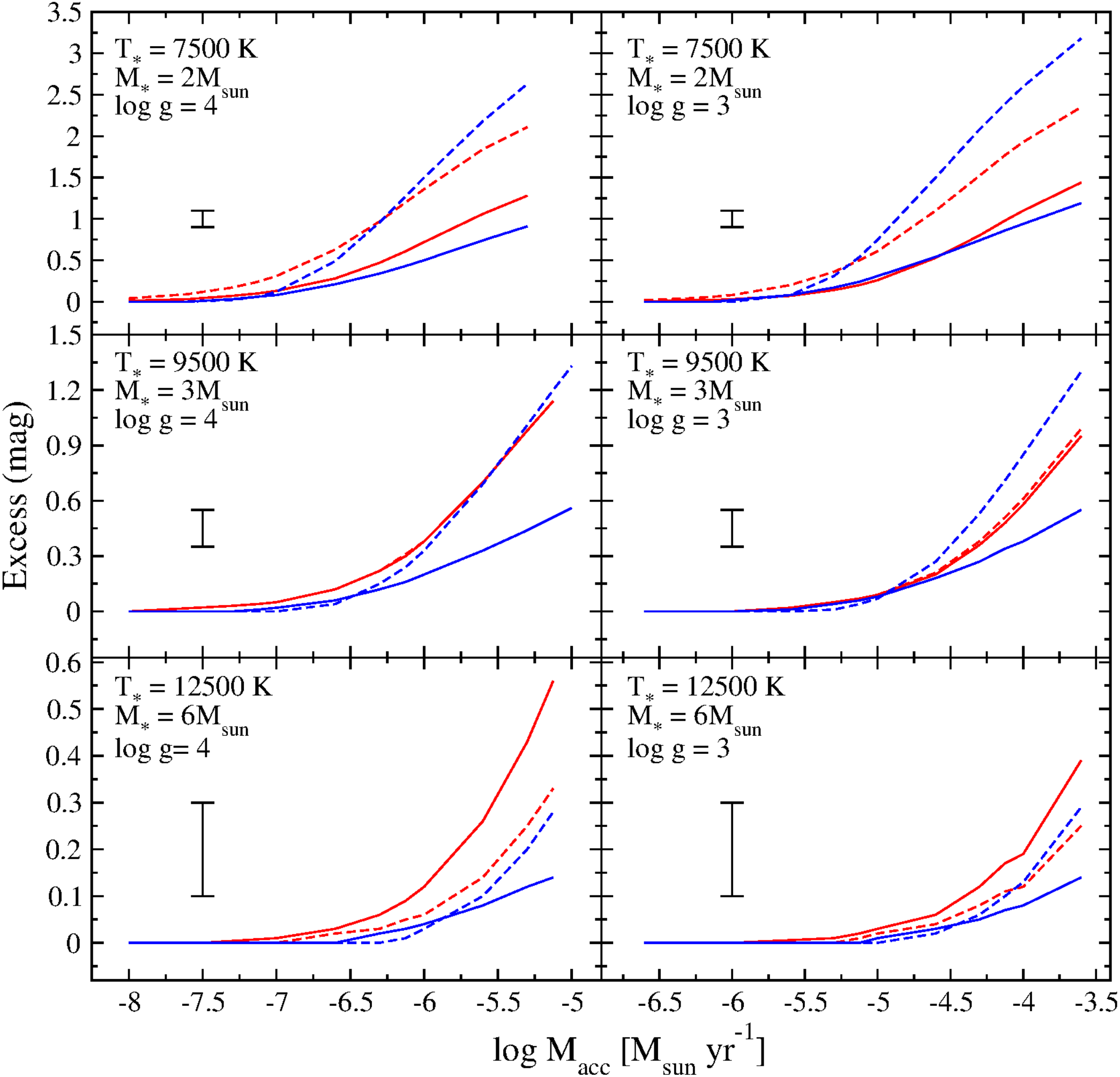}
\caption{Excess predicted from magnetospheric accretion (red lines) and boundary layer (blue lines) shock modeling as a function of the mass accretion rate. The solid lines refer to the Balmer excesses ($\Delta D_{B}$) and the dashed lines to the UV excesses ($\Delta_{UV}$). {The panels show models for a late-type HAe star (top), an early-type HAe star (middle), and a mid-type HBe star (bottom) with the stellar parameters indicated in the legends and surface gravities decreasing from left to right}. All models assume $\cal F$ = 10$^{12}$ erg s$^{-1}$; R$_i$ = 2.5R$_*$ (MA), $\delta_0$ = 0.03, v$\sin$ {\it i} = 100 km s$^{-1}$, and {\it i} = 45$^{\circ}$ (BL). Maximum observational errorbars corresponding to excesses based on broadband photometry are shown.}
\label{Figure:MA_BL_comp}
\end{figure}

Figure~\ref{Figure:MA_BL_comp} compares the excesses predicted by the MA and BL shock models applied to three ``typical'' HAeBes representing a late-type HAe star (top panels), and early-type HAe star (mid-panels), and a mid-type HBe star (bottom panels). Two different values for the gravity at the stellar surface are considered; log g = 4 and 3 for the left and right panels, respectively. Aiming to extract some general conclusions, typical values of $\cal F$ = 10$^{12}$ erg cm$^{-2}$ s$^{-1}$ and R$_i$ = 2.5R$_*$ (for MA), and $\delta_0$ = 0.03, v$\sin$ {\it i} = 100 km s$^{-1}$, and {\it i} = 45$^{\circ}$ (for BL) have been adopted in all cases.

The analysis of Figure~\ref{Figure:MA_BL_comp} provides the following general conclusions:  
\begin{itemize}
\item The relation between the excesses and the accretion rate is strongly dependent on the stellar properties not only for MA (see also Figures 1 and 9 in~\citep{Mendi11b,Fairlamb15} respectively) but also for the BL models. A given excess can correspond to accretion rates different by orders of magnitude, depending on the stellar parameters (T$_*$, M$_*$/R$_*$) of the source. Additional dependencies in the BL model are discussed below. 
\item The differences between predictions from MA and BL become more significant (i.e., above errorbars from broadband photometry) for high excesses/accretion rates. Those differences are generally larger for the Balmer excess than for the UV excess, at least for the space parameters explored here. For instance, the Balmer excess predicted by MA in the mid-left panel of Figure~\ref{Figure:MA_BL_comp} is $\sim$3 times larger than from BL, for $\dot{M}_{\rm acc}$ $\sim$ 3 ~$\times$~ 10$^{-6}$ M$_\odot$ yr$^{-1}$.
\item For the general case analyzed here BL requires higher accretion rates than MA to reproduce a given, large enough Balmer excess. In other words, for relatively large Balmer excesses accretion rates predicted from MA are lower limits to the corresponding from BL, in agreement with the discussion in Section~\ref{sect:BL}.
\item The ratio between the UV and Balmer excesses ($\Delta_{UV}$/$\Delta D_{B}$) predicted by both models tends to differ significantly in most cases. For instance, the ratio predicted in the mid-left panel of Figure~\ref{Figure:MA_BL_comp} is close to unity from MA, while it can reach a factor $\sim$2 from BL. Similarly, for a given star $\Delta_{UV}$/$\Delta D_{B}$ can be $<$ 0 from MA, and $>$ 0 from BL (bottom panels in Figure~\ref{Figure:MA_BL_comp}). This particular difference between predictions represents a good opportunity to observationally compare the two competing models.     
\end{itemize}

\begin{figure}[H]
\centering
 \includegraphics[width=15cm]{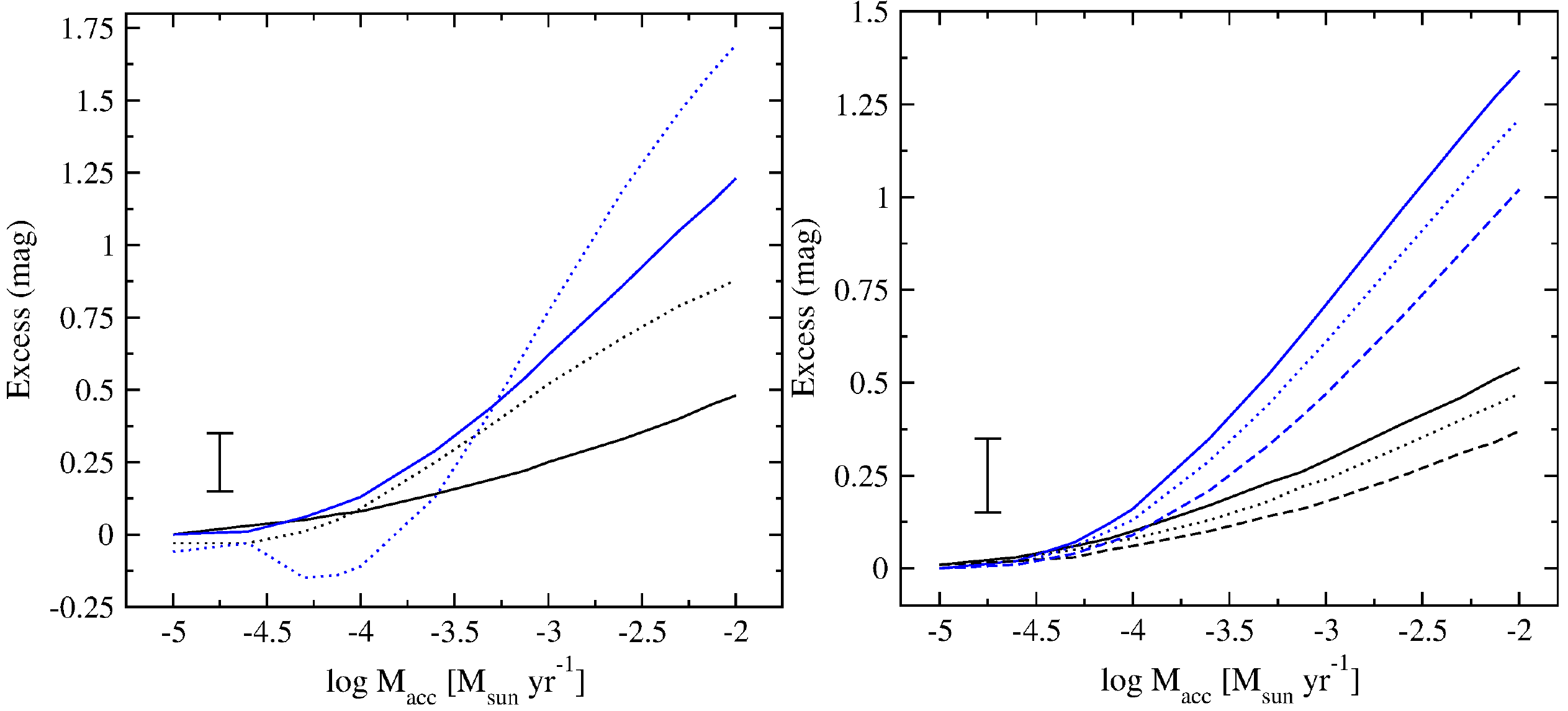}
\caption{Predicted excesses as a function of the accretion rate from BL shock models for a typical HBe star with T$_*$ = 12,500 K, M$_*$ = 6 M$_\odot$, log g = 3, v$\sin$ {\it i} = 100 km s$^{-1}$. The modeled Balmer excesses ($\Delta D_{B}$) and UV excesses ($\Delta_{UV}$) are in black and blue, respectively. Maximum observational errorbars corresponding to excesses based on broadband photometry are shown. Note the different scales in the y-axes. (\textbf{Left panel}) Solid and dotted lines indicate BL sizes given by $\delta_0$ = 0.03 and 0.3, respectively. (\textbf{Right panel}) Different inclinations to the line of sight are shown with solid, dotted, and dashed lines indicating inclinations to the line of sight {\it i} = 30$^{\circ}$, 45$^{\circ}$ and 60$^{\circ}$, respectively.}
\label{Figure:BLdeltai}
\end{figure}

Although the previous analysis indicates that the competing MA and BL scenarios could be eventually distinguished from simultaneous observations at different UV wavelengths, such an analysis does not consider the specific properties of each star-disk system, which, as mentioned before, are crucial to properly reproduce the observations of a given object. Concerning the BL modeling in particular, Figure~\ref{Figure:BLdeltai} shows the influence of varying the size of the BL (left) and the inclination to the line of sight (right). While different inclinations provide excesses that are roughly consistent to each other at least within errors from broadband photometry, the value assumed for the BL size is more critical, and changing it by 1 dex results in significant variations {above photometric errorbars} in the expected excesses both in the Balmer region and at shorter wavelengths. It is recalled that the reason why the large Balmer excesses shown by several HBe stars cannot be explained from MA is that the corresponding filling factors representing the fraction of the stellar surface covered by accretion spots should be above 100$\%$~\citep{Mendi11b,Fairlamb15}, which obviously is not physically possible. In contrast, the large Balmer excesses shown by these HBe stars could be potentially reproduced from BL by changing $\delta_0$ (and thus T$_{\rm BL}$) somewhat arbitrarily, given the lack of observational or theoretical constraints on the BL size. However, as has been shown above, the simultaneous measurements of continuum excesses at different wavelengths limits the possibilities that could be reproduced from BL, potentially constituting a more stringent test of this scenario.

On top of the previous caveats concerning the specifics of each star-disk system, future analysis in the UV must also carefully address the problem of extinction correction. This has not been considered in the previous analysis, but it is critical to properly deal with observational data in that wavelength region, as extinction fundamentally affects the excesses that want to be reproduced.

The Hubble Space Telescope’s Ultraviolet Legacy Library of Young Stars as Essential Standards (ULLYSES)\footnote{\url{http://www.stsci.edu/stsci-research/research-topics-and-programs/ullyses}.} will soon observe a large sample of young stars in the UV probing the lowest and the highest stellar temperatures (K-M and early O-B stars). Although ULLYSES will not cover the intermediate-mass regime, the resulting dataset in combination with related ground-based efforts will surely provide valuable strategies to best deal with related archival data of HAeBe stars. In the longer term, $>$100 h with the future World Space Observatory\footnote{\url{http://www.wso-uv.es/index.php?id=33}.} will in principle be devoted to observe HAeBes in the UV and tackle the problem of the best suitable accretion scenario in these stars.

\section{Concluding Remarks}
\label{sect:summary}
In this work our estimates of the mass accretion rates in HAeBe stars have been reviewed. Guided by previous works devoted to the better known low-mass regime, we have seen that MA also seems to be a valid scenario that explains the properties of the late-type HAes and is capable of providing accretion rates $\sim$ an order of magnitude larger than for $\sim$1 M$_{\odot}$ CTTs. Still, future work is necessary to reach a consensus on which accretion scenario is best suited for the HAeBe regime. Two main lines of research have been suggested to clarify this open issue. First, it is necessary to reduce the detection limits involved in the determination of magnetic fields in HAeBes, which currently are of the same order than the minimum magnetic fields required to drive accretion magnetospherically in these sources. Secondly, more efforts should be devoted to the theoretical and observational developments of alternative accretion scenarios. Although it has been shown from {simple} accretion shock {models} that the UV region could help to disentangle between MA and the competing BL scenario, {proper modeling of the BL region is still pending. Indeed, radiative transfer models able to predict the shape and strength of emission lines from BL would be particularly useful}. Alternatively, other possible accretion paradigms different than MA and BL could work in HAeBes (see e.g.,~\citep{Takasao18}).

Even assuming that MA can be extended from the CTTs to the late-type HAes, MA shock modeling is not capable of reproducing the Balmer excesses of several early-type HBe stars, which require a different approach~\citep{Mendi11b,Fairlamb15}. If accretion in these sources occurs through a BL, the analysis presented in this work suggests that they would be accreting at very high rates, larger than inferred from the MA-based correlations with the emission lines or the stellar luminosities. For instance, the large photometric Balmer excesses of HBe stars that cannot be explained from MA are $>$0.5 magnitudes and could reach $>$1 magnitudes for some sources~\citep{Mendi11b}, in principle corresponding to rough accretion rates $>$10$^{-3}$ M$_{\odot}$ yr$^{-1}$ using large enough BL sizes (Figure~\ref{Figure:BLdeltai}, left). Such huge accretion rates are indeed higher than in principle expected for HBe sources (e.g.,~\citep{Beltran16}, and references therein), which by definition have already dissipated their natal envelopes and are very close to the MS. Although this review has been focused on accretion, alternative physical processes that do not invoke infalling material may also be capable of explaining the observed excesses in HBes, and perhaps several other observational differences between these and lower-mass stars. In particular, there is evidence that photoevaporation plays a more important role in HBes than in HAes (see e.g.,~\citep{Alonsoalbi09,Wolff11,Guzman20}, and references therein). In turn, photoevaporated disks should show relatively small accretion rates according to models (e.g.,~\citep{Clarke01,Gorti09}, and references therein), again suggesting that $>$ 10$^{-3}$ M$_{\odot}$ yr$^{-1}$ may be unrealistically large for HBes. Therefore, it is worth exploring the possibility that the observed Balmer and UV excesses in HBes are not explained mainly in terms of accreting material shocking onto the star, as for HAes and CTTs, but perhaps in terms of photoevaporative outflows shocking onto the disks.

\vspace{6pt} 

\funding{The author is funded by a ``Talento'' Fellowship (2016-T1/TIC-1890; Government of Comunidad Aut\'onoma de Madrid, Spain). This research has also been partially funded by the Spanish State Research Agency (AEI) Project No. ESP2017-87676-C5-1-R and No. MDM-2017-0737 Unidad de Excelencia “María de Maeztu”- Centro de Astrobiolog\'{\i}a (CSIC-INTA).}

\acknowledgments{The author sincerely acknowledges James Muzerolle, Antonella Natta, Rene Oudmaijer, and Nuria Calvet for their useful comments based on a preliminary version. The author also acknowledges the two referees for their positive reports and their suggestions, which have served to improve the~manuscript.}

\conflictsofinterest{The author declares no conflict of interest.}

\newpage
\abbreviations{The following abbreviations are used in this manuscript:\\

\noindent 
\begin{tabular}{@{}ll}
BL & Boundary Layer\\
CTT & Classical T Tauri\\
HAe & Herbig Ae\\
HAeBe & Herbig Ae/Be\\
HBe & Herbig Be\\
IMTT & Intermediate-Mass T Tauri\\
IR & Infrared\\
IUE & International Ultraviolet Explorer\\
MA & Magnetospheric Accretion\\
MS & Main Sequence\\
MYSO & Massive Young Stellar Object\\
PMS & Pre-Main Sequence\\
SFR & Star Formation Rate\\
TT & T Tauri\\
ULLYSES & Ultraviolet Legacy Library of Young Stars as Essential Standards\\
UV & Ultraviolet\\
WTT & Weak T Tauri\\
\end{tabular}}


\reftitle{References}
\label{refs}




\end{document}